\documentclass[aps,prb,twocolumn]{revtex4}
\usepackage{tabularx,graphicx}
\def\gtwid{\,{\raise.3ex\hbox{$>$\kern-.75em\lower1ex\hbox{$\sim$}}}\,}
\def\ltwid{\,{\raise.3ex\hbox{$<$\kern-.75em\lower1ex\hbox{$\sim$}}}\,}

\begin{document}
\title{Granular systems in the Coulomb blockade regime} 

\author{D. P. Arovas$^1$, F. Guinea$^2$, C. P. Herrero$^2$,
P. San Jos\'e$^2$}
\affiliation{
$^1$ Department of Physics, University of California at San Diego,
 La Jolla CA 92093 \\$^2$
Instituto de Ciencia de Materiales de Madrid,
CSIC, Cantoblanco, E-28049 Madrid, Spain.}
\date{\today}
\begin{abstract}
Disordered granular systems, at temperatures where charging effects are
important, are studied, by means of an effective medium approximation.
The intragrain charging energy leads to insulating behavior at low
temperatures, with a well defined Coulomb gap. Non equilibrium effects
can give rise to a zero temperature transition between a metallic,
gapless phase, and an insulating phase. 
\end{abstract}
\maketitle

\section{Introduction}
Charging effects modify the
transport properties of granular 
metals at low temperatures\cite{GZ68,LJ69,S92}. In magnetic systems,
the low temperature magnetoresistance can also be modified
by charging 
effects\cite{Betal98,Cetal98,Metal98,ZW99,Getal00,Fetal02}.

A successful scheme used to analyze Coulomb blockade in granular 
systems is to reduce the system to the study of an effective
single junction, as in the pioneering
study done in Ref. \onlinecite{HA76}. This assumes that
each junction has an activated conductance, which is 
approximately correct when the high temperature conductance
of the junction  $g$, is low in units of $e^2 / \hbar$. 
A single junction in the Coulomb blockade regime
has been extensively
studied in recent years\cite{CB}. Modern techniques allow us
to study the behavior of a single junction for arbitrary values
of the conductance and offset charges (see below).
 
The current approach 
describes a single junction 
in terms of two collective variables, 
the charge, $Q$, and its conjugate phase variable,
$\phi$. The electronic 
degrees of freedom are integrated out\cite{H83,BMS83}.
This procedure is motivated by a similar approach used in
the study of small Josephson junctions\cite{AES82}. 
The present tools allow us to study junctions with high values 
of the conductance,
$g \gg e^2 / \hbar$, and to consider the effect of gate voltages,
and of degeneracies between different charge states.
The model can be extended to study non equilibrium processes
associated to the tunneling events\cite{UK90,UG91}, which
can be considered as the leading corrections in an expansion
in $g^{-1}$, where $g$ is the intragrain conductance\cite{Betal00,BMM97}.

In the following, we will extend these standard methods
used in the study of Coulomb blockade in single junctions to
the analysis of granular materials. In order to do so,
we will use an effective
medium theory. This approach can be viewed 
as an extension
to mesoscopic systems of the
Dynamical Mean Field Theory used in the study of
the microscopic properties of strongly correlated 
systems\cite{Getal96}. The model and
method of calculation will be presented in the next
section. Section III discusses results obtained using
analytical large-N  techniques which describe well the properties
of single junctions\cite{R97,AD02}. Then, we show
numerical results obtained using Monte Carlo (MC) methods
previously applied to the study of single junctions\cite{HSZ99}.
Section IV contains the main conclusions of our work.
 
\section{The model}
\subsection{Effective medium theory.}
We will consider a system made of metallic grains. We assume that the
separation of the electronic states within each grain, $\delta_i$, is much
smaller than the charging energies, $E_{\rm c}^i$, or the temperature,
$k_{\rm B}T$.
The grains are connected by a large number of channels
with small transmission coefficients, so that a perturbative
treatment of the coupling induced by each individual channel
is possible. Then, the electronic degrees of freedom can be
integrated out\cite{AES82,H83,BMS83,SZ90}, and the system 
is described in terms of the charge at each grain, $Q_i$, and
its conjugated phase variable, $\phi_i$, where
$[ Q_i , \phi_i ] = i e$, and $e$ is the unit of 
charge. The effective action which
describes the charge effects in the system is, at zero temperature:
\begin{widetext}
\begin{equation}
{\cal S} = {1\over 2e^2}\sum_{i,j} \int\limits_0^\beta
d\tau\, C_{ij}(\partial_\tau\phi_i) (\partial_\tau\phi_j)
+\sum_{i<j}\alpha_{ij}\int\limits_0^\beta\!\! d\tau\!\!\int\limits_0^\beta \!\!d\tau'
K(\tau-\tau')\bigg\{1-\cos\Big[\phi_i(\tau)-\phi_j(\tau)-
\phi_i(\tau')+\phi_j(\tau')\Big]\bigg\}\ ,
\label{action_eff}
\end{equation}
\end{widetext}
where $K(\tau)=[\pi T {\rm csc}(\pi T\tau)]^2$ (we henceforth set
$k_{\rm B}\equiv 1$).  We further assume that the
capacitance matrix is dominated by its diagonal entries, and identify
the charging energy of the $i^{\rm th}$ grain as $E_{\rm c}^i=e^2/2C_{ii}$.
The coupling between grains $i$ and $j$ is $\alpha_{ij}=2|t_{ij}|^2\rho_i\rho_j$,
where $t_{ij}$ is the hopping matrix element between the grains (assumed 
independent of momenta), and $\rho_{i,j}$ are the respective densities of states.
One can also write $\alpha_{ij}=R_{\rm K}/4\pi^2 R_{ij}$, where $R_{\rm K}=
h/e^2=25,813\,{\rm \Omega}$ is the quantum of resistance and $R_{ij}$ is the
high-temperature value of the intergrain resistance.

The above model possesses a global ${\rm O}(2)$ rotational symmetry,
which may be generalized to an ${\rm SU}(N)$ symmetry as follows.
On each site, we replace the unimodular complex scalar $z_i\equiv e^{i\phi_i}$
with an $N$-component unimodular complex vector with components $z_{i\mu}$
($\mu=1,\ldots,N$).  The action for the ${\rm SU}(N)$ model is then
\begin{widetext}
\begin{equation}
{\cal S}=\sum_i {1\over 4 E^i_{\rm c}}\int\limits_0^\beta d\tau\,
(\partial_\tau z^*_{i\mu})(\partial_\tau z_{i\mu})
+\sum_{i<j}\alpha_{ij}\int\limits_0^\beta\!\! d\tau\!\!\int\limits_0^\beta\!\! d\tau'
K(\tau-\tau')\,\bigg\{1-z^*_{i\mu}(\tau)z_{j\mu}(\tau)z_{i\nu}(\tau')
z^*_{j\nu}(\tau')\bigg\}
\end{equation}
\end{widetext}
We now assume that each grain is coupled to a large number, $M$, of other
grains, and define the mean field correlation function $G_i(\tau-\tau')$ as
\begin{equation}
{M\over N}\,G_i(\tau-\tau')\,\delta_{\mu\nu}\equiv
{1\over {\bar\alpha}}\Big\langle\sum'_j \alpha_{ij}\, z_{j\mu}(\tau)
z^*_{j\nu}(\tau')\Big\rangle\ ,
\end{equation}
with ${\bar\alpha}\equiv M^{-1} \langle\sum_j\alpha_{ij}\rangle$ the average
coupling of a grain to its neighbors.  Note that $G(0)=1$.  This allows us to
write the action as a sum over contributions from individual grains, with the
effective self-interaction term
\begin{equation}
{\cal S}^{\rm int}_i={\tilde\alpha}\int\limits_0^\beta\!\! d\tau\!\!
\int\limits_0^\beta\!\! d\tau'
K(\tau-\tau')\,\Big\{1-G_i(\tau-\tau')z^*_{i\mu}(\tau)z_{i\mu}(\tau')\Big\}
\end{equation}
with ${\tilde\alpha}\equiv M{\bar\alpha}/N$.
Thus, the properties of the system can be reduced to those
of a single junction embedded in an effective environment
whose properties are to be determined self consistently.

\subsection{Offset charges and non equilibrium effects.}
We can extend the model discussed so far to situations when
there are offset charges, $q_{i}$, at each grain.
These offset charges lead to additional phases in the
contributions of paths of $\phi ( \tau)$ with non zero
winding numbers. A path of winding number $m$ acquires
a phase $\chi_i = 2 \pi m q_i$. This phase has to be 
added to the effective action, Eq.(\ref{action_eff}).

The same scheme can also be used to include the effect
of non equilibrium processes associated to  
intergrain tunneling\cite{UK90,UG91}. The electron which tunnels induces
an inhomogeneous electrostatic potential, which modifies
the electronic levels within the grain. This effect can
be analyzed using similar techniques to those used in relation to
the sudden ejection of a core electron in a metal\cite{M91}, and it
leads to a modification of the kernel in the term which describes
the tunneling in the action in Eq.(\ref{action_eff}).
The decay of the interaction $K(\tau-\tau')$ at long times becomes
$|\tau - \tau'|^{2 - \epsilon}$, where $\epsilon$ is
a dimensionless parameter, which can be expressed in terms
of the phaseshifts induced at the Fermi level by the sudden
switching of the external potential. A similar effect is expected
if the tunneling takes place between quasi one-dimensional 
systems\cite{KF92}.

In fact, for each pair of grains one should replace the kernel $K(\tau)$ with
\begin{equation}
K_{ij}(\tau)=(E^i_{\rm c})^{\epsilon_i}\,(E^j_{\rm c})^{\epsilon_j}\,
[\pi T\csc(\pi T\tau)]^{2-\epsilon_i-\epsilon_j} \ .  
\end{equation}
For the sake of simplicity, we will assume
\begin{equation}
K(\tau)={\overline{E_{\rm c}}}^\epsilon [\pi T\csc(\pi T\tau)]^{2-\epsilon}
\label{kernel}
\end{equation}
for all pairs, where $\overline{E_{\rm c}}$ is the average charging energy.

\subsection{Monte Carlo procedure}
In our effective-medium scheme, the grand partition function for a generic 
grain $i$ with a given offset charge $q_i$ can be written in terms
of the phase $\phi_i$, as \cite{SZ90}:
\begin{equation}
  Z(q_i) = \sum_{m=-\infty}^{\infty} \exp(i \chi_i) \,
       \int {\cal D} \phi_i \, \exp \left( - {\cal S}_{\rm eff}[\phi_i] \right)
  \hspace{.2cm} ,
  \label{part}
\end{equation}
where the paths $\phi_i(\tau)$ satisfy in sector $m$ the
boundary condition $\phi_i(\beta) = \phi_i(0) + 2 \pi m$.
The effective action takes the form:
\begin{widetext}
\begin{equation}
{\cal S}_{\rm eff}[\phi] = \frac{1}{4 \overline{E_{\rm c}}} \int_0^{\beta} d \tau
\left( \frac{\partial \phi}{\partial \tau} \right)^2 + \
 {\tilde\alpha} \int_0^{\beta} d\tau \int_0^{\beta} d\tau' K(\tau-\tau') \
\{ 1 - \cos [\phi(\tau)-\phi(\tau')] \ G(\tau-\tau') \}    
\label{action_eff2}
\end{equation}
\end{widetext}
with the kernnel $K(\tau)$ given in Eq. (\ref{kernel})
It describes low-energy processes below an upper cutoff on the
order of the unscreened average charging energy, $\overline{E_{\rm c}}$.
Here, the correlation function $G$ is given by 
\begin{equation}
G(\tau-\tau') = \langle \cos [\phi(\tau) - \phi(\tau')] \rangle \ ,
\label{correl}
\end{equation}
where the average value is calculated with the partition function in
Eq. (\ref{part}).
Note that the action in Eq. (\ref{action_eff2}) is similar to that corresponding
to a single tunnel junction with dimensionless conductance ${\tilde\alpha}$
and charging energy $\overline{E_{\rm c}}$, the only difference being the presence of
the correlator $G$ in the tunneling part of the effective action of the
granular system.
          
Different sectors (winding numbers $m$) in the partition function $Z(q_i)$ 
have been sampled by
path-integral Monte Carlo for temperatures down to $T = \overline{E_{\rm c}}/50$.
MC simulations have been carried out by the usual discretization of
quantum paths into $N$ (Trotter number) imaginary-time slices \cite{S93}.
To keep roughly the same precision in the calculated
quantities, as the temperature is reduced, the number of time slices
has to be increased as $1/T$. We have taken $N = 4 \beta \overline{E_{\rm c}}$, and
thus, we employed an imaginary-time slice
$\Delta \tau = \beta / N = 1 / (4 \overline{E_{\rm c}}) $.
Details on this kind of MC simulations were given in earlier publications
\cite{HSZ99,Betal00} and will not be repeated here.
 
>From a computational point of view, the main difference between
the present procedure and that used in earlier works \cite{HSZ99,Betal00}
is the requirement of self-consistency for the correlator
$G(\tau - \tau')$, as expressed by Eqs. (\ref{action_eff2}) and (\ref{correl}).
To obtain this self-consistency, we employed an iterative procedure that
led to convergence of the correlator in the region of parameters studied
here. For a given set of parameters
($T$, ${\tilde\alpha}$, $\epsilon$, and $q_i$), we
begin the simulations by assuming a correlator $G_0(\tau - \tau') = 1$.
The correlator $G_1$ obtained by introducing $G_0$ in the
effective action Eq.(\ref{action_eff2}) is then
employed in a new iteration to obtain $G_2$, and so on.
In our simulations, each iteration consisted of $2 \times 10^4$ MC steps
(in a MC step all path coordinates and winding number are updated).
By using this procedure, the correlator converges after a few iterations.
In general, the number of iterations required for convergence increases 
as the temperature is lowered, 
but in fact 10 iterations were enough to find convergence
in all cases presented here. We have checked that employing different
starting correlation functions $G_0(\tau - \tau')$ in our iterative 
procedure does not change the results.
                   
The conductance $g(T)$ has been obtained from the (self-consistent)
correlator $G$ by using the expression \cite{Betal00}:
\begin{equation}
g(T) = g_0 \left( \frac{\overline{E_{\rm c}}}{T} \right)^{\epsilon}
 G \left( \beta/2 \right) \ ,
\end{equation}
where $g_0$ is the high-temperature conductance.

\section{Results}

\subsection{Large-$N$ Theory}
We now address the self-consistent single grain theory within the large-$N$
approximation, where $N$ is the number of components of the field $z_\mu$.
The unimodularity constraint $\sum_{\mu=1}^N |z_{\mu}|^2=1$ can be imposed
with a real Lagrange multiplier field, $\lambda(\tau)$:  
\begin{eqnarray}
{\cal S}&=&\int\limits_0^\beta \left\{{1\over 4E_{\rm c}}|\partial_\tau z_\mu|^2
+\lambda (|z_\mu|^2-1)\right\}\\
&&+{\tilde\alpha}\int\limits_0^\beta\!\! d\tau\!\!
\int\limits_0^\beta\!\! d\tau'
K(\tau-\tau')\,\left\{1-G(\tau-\tau')z^*_\mu(\tau)z_\mu(\tau')\right\}\nonumber
\end{eqnarray}
where $G(\tau-\tau')=\langle z^*_\mu(\tau) z_\mu(\tau')\rangle$ is the
self-consistent field due to neighboring sites.  The partition function is then obtained by
integrating $\exp(-{\cal S})$ freely over the complex fields $\{z_\mu(\tau),
{\bar z}_\mu(\tau),\lambda(\tau)\}$.  In the $N\to\infty$ limit, the
saddle point of the functional integral is dominated by configurations in
which $\lambda(\tau)$ is constant.  The action, up to a harmless constant, is
\begin{equation}
{\cal S}=\sum_{\omega_n} \left(\frac{1}{4}\omega_n^2 + \lambda
+ {\tilde\alpha} [{\hat Q}(0)-{\hat Q}(\omega_n)]\right)
|{\hat z}_\mu(\omega_n)|^2 -\lambda\ ,
\end{equation}
where we now express all energies in units of $E_{\rm c}$.  We define 
$L\equiv\beta E_{\rm c}$; the bosonic Matsubara
frequencies are $\omega_n=2\pi n/L$.  The function ${\hat Q}(\omega_n)$
is the Fourier transform of $Q(s)\equiv K(s) G(s)$, where
$s\equiv E_{\rm c}\tau$ is the dimensionless imaginary time, and
\begin{equation}
K(s)=\left[{\pi\over L}{\rm csc}\left({\pi |s|\over L}\right)
\right]^{2-\epsilon}\ .
\end{equation}
Thus,
\begin{equation}
{\hat Q}(0)-{\hat Q}(\omega_n)=\int\limits_0^L\!ds\,
(1-\cos(\omega_n s))\,K(s) G(s)\ .
\end{equation}
The correlation function $G(s)$ is given by
\begin{equation}
G(s)={N\over L}\sum_{\omega_n}{e^{i\omega_n s}\over
\frac{1}{4}\omega_n^2 +\lambda + {\tilde\alpha}[{\hat Q}(0)
-{\hat Q}(\omega_n)]},
\end{equation}
and the fixed length constraint is $G(0)=1$.

\begin{figure}
\includegraphics[width=18cm]{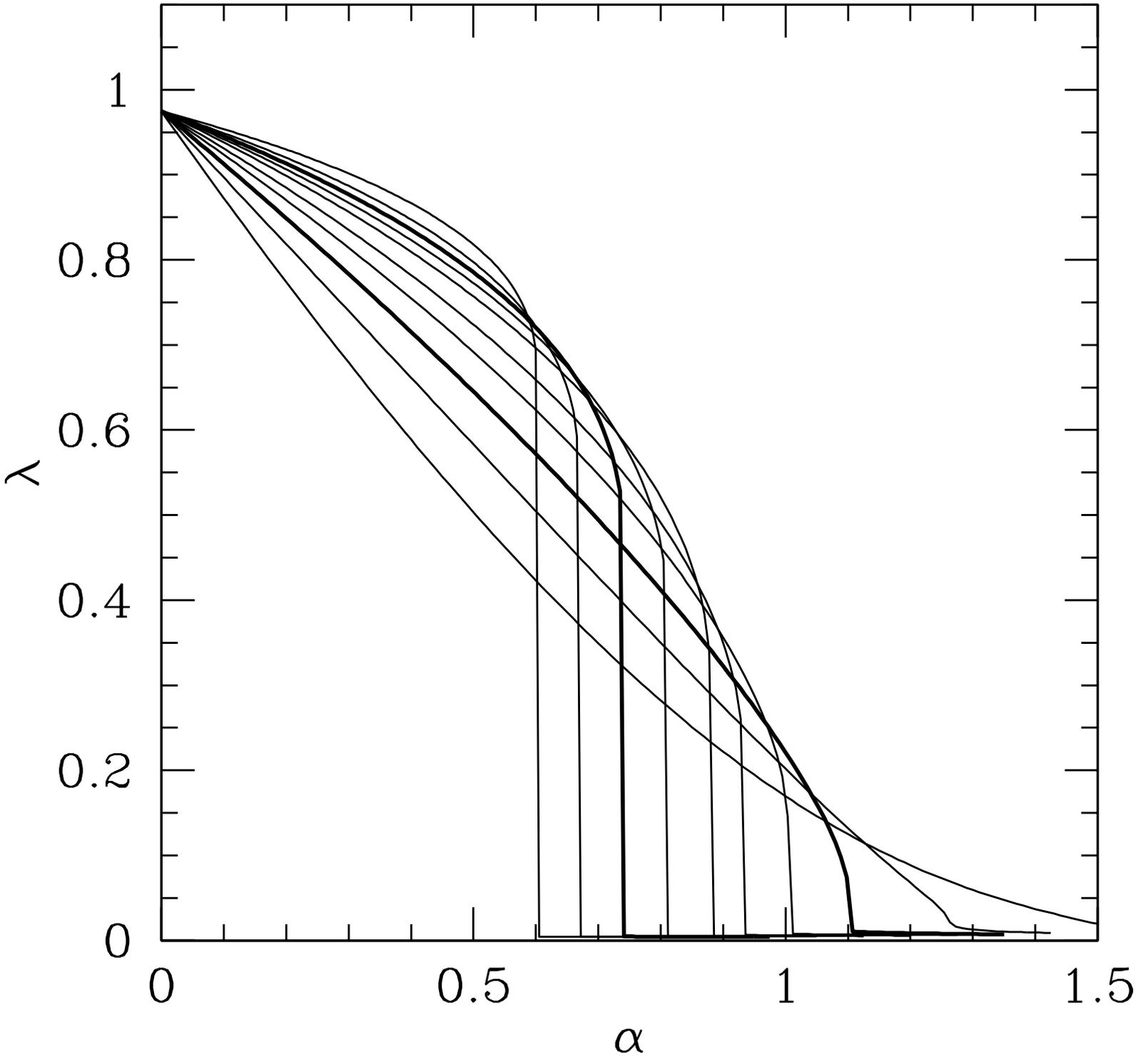}
\caption{\label{lvsa} Solution to the large-$N$ theory for $L=128$ and $\epsilon=0.1,
0.3,0.5,\ldots,1.9$.  ${\tilde\alpha}_{\rm c}$ decreases with increasing $\epsilon$.  Thick
curves are for $\epsilon=0.5$ and $\epsilon=1.5$.}
\end{figure}

We iterate the mean field equations at finite temperature to obtain
$\lambda({\tilde\alpha},L)$ in the following manner.  First, given an
initial guess for the set $\{{\hat Q}(\omega_n)\}$, we find $\lambda$
such that $G(0)=1$.  We then recompute $Q(s)=K(s)G(s)$ and a new set
of Fourier components $\{{\hat Q}'(\omega_n)\}$.  Iterating until self-consistency,
we obtain the results shown in Fig. \ref{lvsa}.  The self-consistent solutions reveal
a discontinuous jump in $\lambda$ at ${\tilde\alpha}_{\rm c}$, {\it i.e.\/} the transition
apparently is first order.

However, it is hardly clear that our iteration scheme should converge to the
proper ({\it i.e.\/} lowest free energy) self-consistent solution.  If the transition is
second order, we can make progress by assuming, at criticality
($L=\infty$, ${\tilde\alpha}={\tilde\alpha}_{\rm c}$), that
\begin{equation}
G(s)=N\int\limits_0^\infty {d\omega\over\pi}\,{\cos(\omega s)\over
\frac{1}{4}\omega^2 + {\tilde\alpha}_{\rm c}[{\hat Q}(0)-{\hat Q}(\omega)]}\ ,
\label{gcrit}
\end{equation}
with $Q(s)=G(s)/|s|^{2-\epsilon}$.  Since $G(0)=1$, we have that $Q(s)\simeq
|s|^{\epsilon-2}$ as $s\to 0$.  For $s$ large, we assume $Q(s)\simeq A |s|^{\delta-2}$
and solve for $A$ and $\delta$ by ignoring the first term in the denominator of
Eq. (\ref{gcrit}).  We find
\begin{equation}
A=\left({N\over 4\pi\alpha_{\rm c}}(2-\epsilon){\rm ctn}(\pi\epsilon/4)\right)^{1/2}
\end{equation}
and $\delta=\frac{1}{2}\epsilon$.  Accordingly, we adopt a trial solution for $Q(s)$ which
interpolates between the small and large $s$ limits:
\begin{equation}
Q(s)={e^{-\eta |s|}\over |s|^{2-\epsilon}}+{A(1-e^{-\eta |s|})\over |s|^{2-\frac{1}{2}\epsilon}}\ .
\end{equation}
with $\eta$ a variational parameter.

\begin{figure}
\includegraphics[width=18cm]{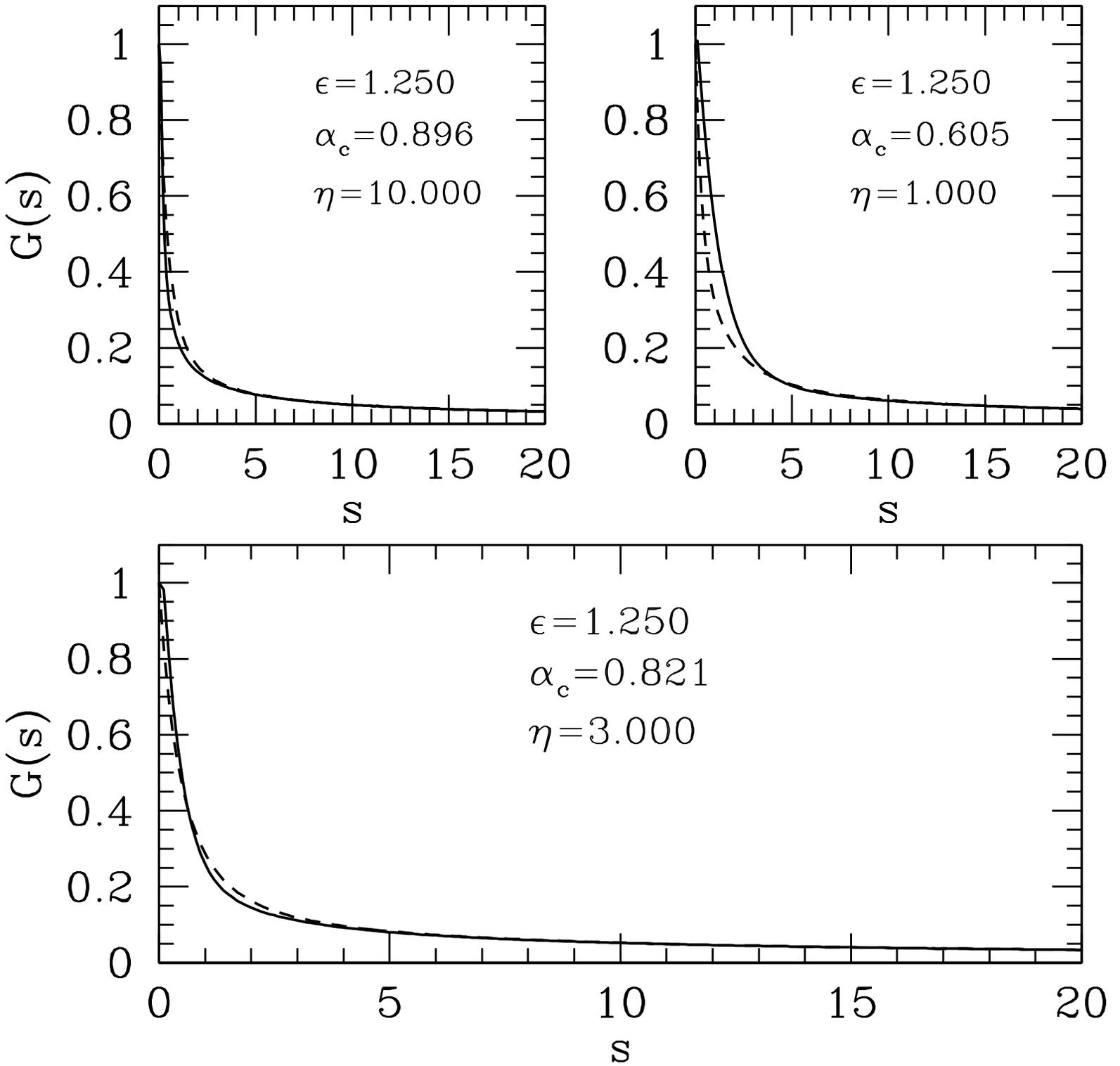}
\caption{\label{gvss}Correlation function $G(s)$ at criticality {\it versus\/}
dimensionless imaginary time $s$.   Solid line is input $G(s)=e^{-\eta |s|}
+A (1-e^{-\eta |s|}) |s|^{-\frac{1}{2}\epsilon}$; dashed line is obtained from
Eq. \ref {gcrit}.  Here $\eta=3.00$ provides the best fit of the three curves.}
\end{figure}

Given this trial form for $Q(s)$, we solve $G(0)=1$ to obtain ${\tilde\alpha}_{\rm c}(\epsilon,\eta)$.
We then check for self-consistency, comparing $G(s)$ obtained from Eq. (\ref{gcrit}) with
$Q(s)\,|s|^{2-\epsilon}$.  We find (i) that ${\tilde\alpha}_{\rm c}$ is insensitive to $\eta$
for $\epsilon \ltwid 0.25$, but increasingly sensitive for larger values, and (ii) that there is an
optimal value for $\eta$, in order that the two definitions of $G(s)$ agree best (see Fig. 
\ref{gvss}).

\begin{figure}
\includegraphics[width=18cm]{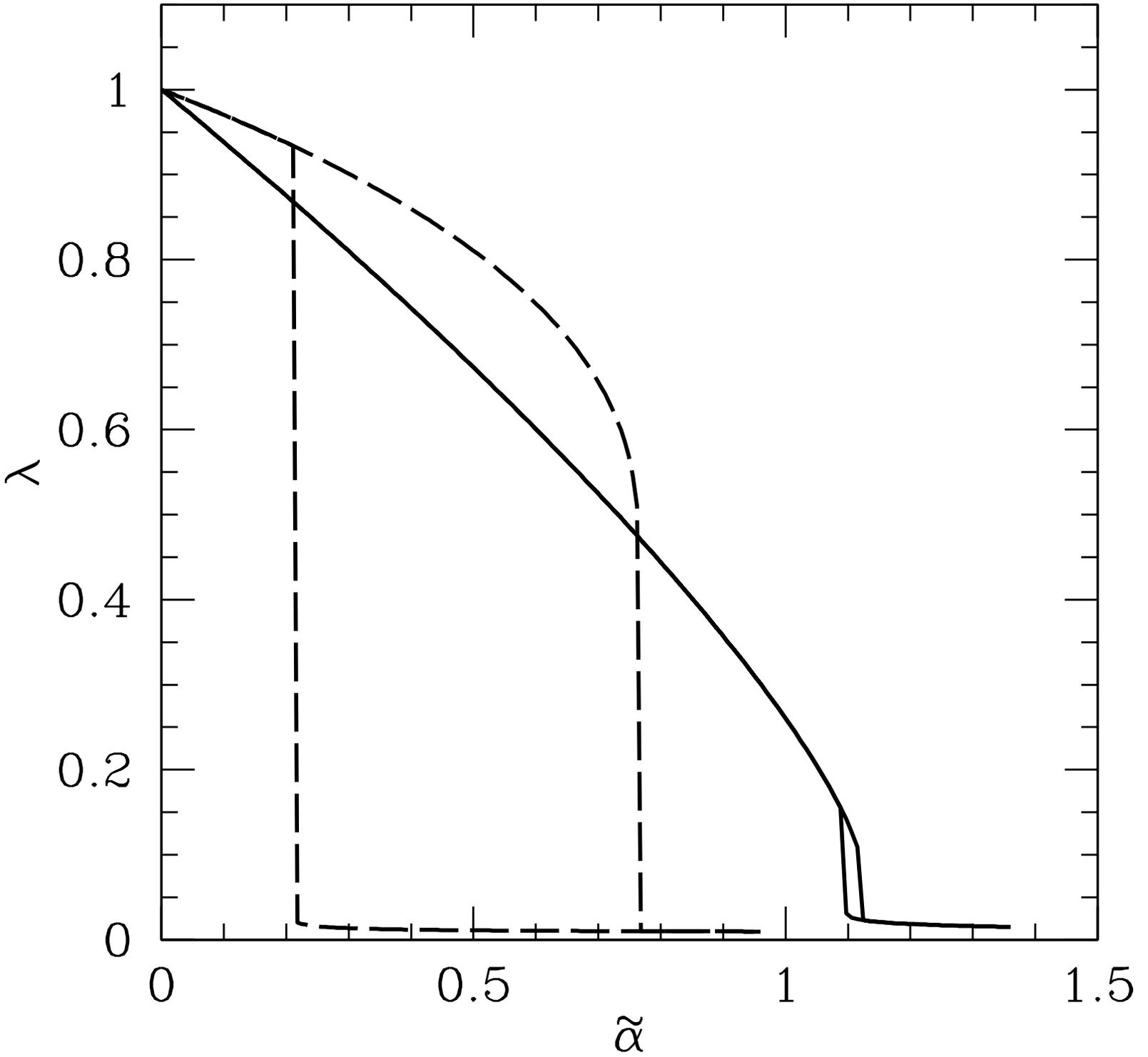}
\caption{\label{hyst}Hysteretic behavior in the self-consistent solution for $L=128$.
Solid curves are for $\epsilon=0.50$; dashed curves are for $\epsilon=1.50$. }
\end{figure}

\begin{figure}
\includegraphics[width=18cm]{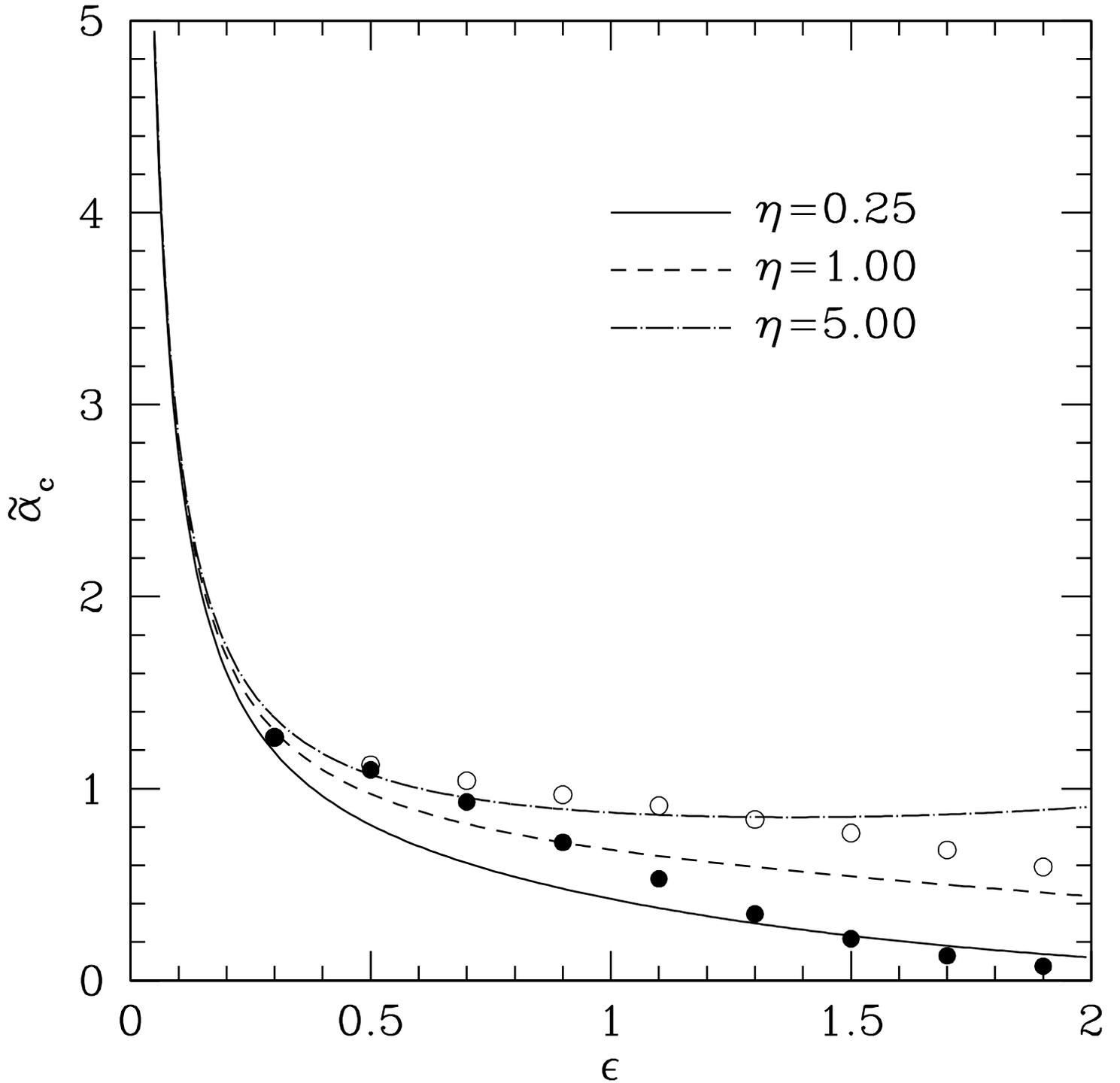}
\caption{\label{avse}${\tilde\alpha}_{\rm c}$ {\it versus\/} $\epsilon$ for three different
values of $\eta$ (curves).  Open and closed circles denote maximum and minimum
values of ${\tilde\alpha}_{\rm c}$ inferred from iterative solution to the self-consistent 
equations with $L=128$. }
\end{figure}

Returning to the first order behavior found in the iterated self-consistent
solution, we find that this behavior persists for $\epsilon\gtwid 0.30$.  Furthermore,
it is hysteretic.  The solution shown in Fig. \ref{lvsa} was obtained
by sweeping ${\tilde\alpha}$ up from ${\tilde\alpha}=0$.  When a self-consistent
solution was obtained, the set $\{ {\hat Q}(\omega_n)\}$ was chosen as the initial
conditions for iteration for the next value of ${\tilde\alpha}$.  
When we sweep {\it both\/}
up and down, however, we find the solution is hysteretic, as shown in Fig. \ref{hyst}.
In Fig. \ref{avse} we compare ${\tilde\alpha}_{\rm c}$ obtained assuming a second
order transition, as discussed above, with the upper and lower critical values obtained
by iteration of the self-consistent equations (with $L=128$).

\subsection{Monte Carlo simulations}

We now turn to the results of our path-integral Monte Carlo simulations for
the conductance of the granular system.

\begin{figure}
\includegraphics[width=18cm]{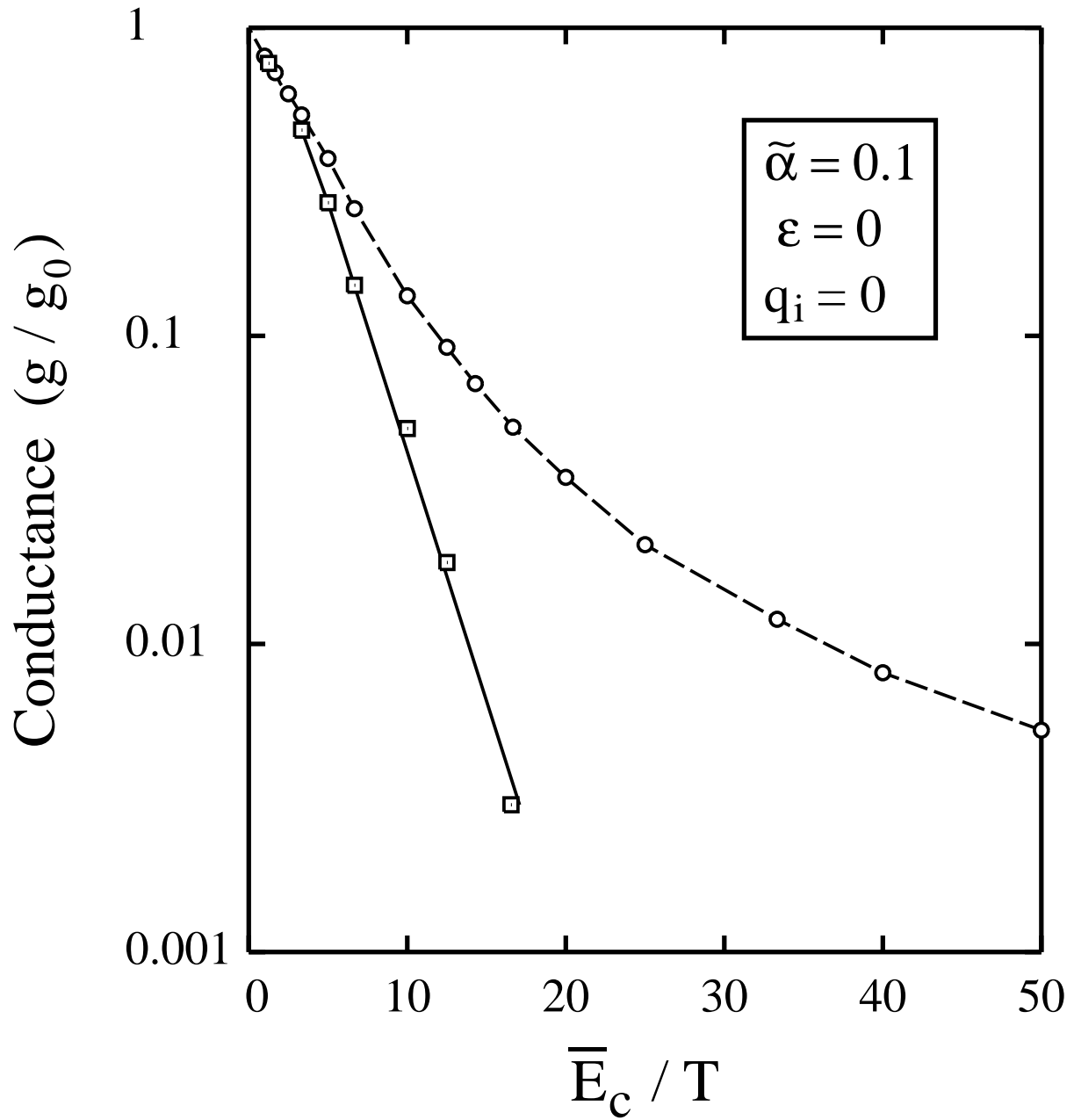}
\caption{\label{gt}
  Temperature dependence of the normalized conductance $g/g_0$, as derived from
path-integral Monte Carlo simulations for a granular system (squares) and for
a single tunnel junction (circles) with the same parameters: ${\tilde\alpha} = 0.1$,
$\epsilon = 0$, and $q_i = 0$ (no offset charge). Error bars are on the order
of the symbol size.  }
\end{figure}

In Fig. \ref{gt} we present the temperature dependence of the conductance $g(T)$,
as derived from our self-consistent correlation function $G$ (squares)
for ${\tilde\alpha} = 0.1$ and $\epsilon = 0$, corresponding to a Coulomb-blockade 
regime.
For comparison, we also show MC results found for the same set of parameters for
a single junction (circles), for which one finds at low $T$ an algebraic dependence
of the conductance in the form $g(T) \sim T^2$, as expected for co-tunneling
processes. These processes do not show up in the granular system, and in this case 
$g(T)$ decreases exponentially as the temperature is lowered.

\begin{figure}
\includegraphics[width=18cm]{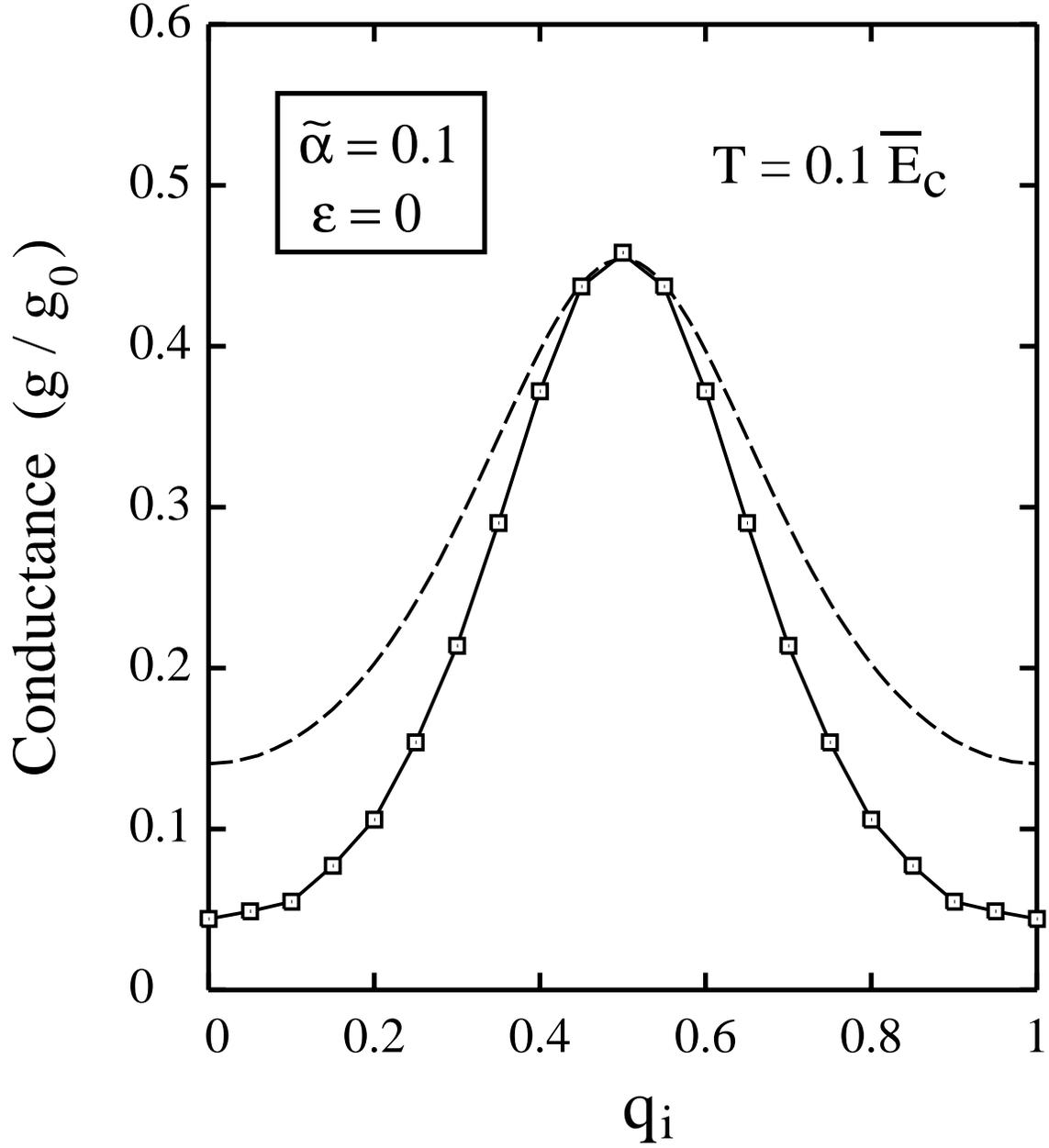}
\caption{\label{gqi}
 Normalized conductance ($g/g_0$) vs. offset charge $q_i$, as derived from
Monte Carlo simulations for a granular system (squares) and for a single tunnel
junction (dashed line) with the same parameters: ${\tilde\alpha} = 0.1$,
$\epsilon = 0$, and $T = \overline{E_{\rm c}} / 10$. }
\end{figure}

Our Monte Carlo procedure allows us to study also the dependence of the
conductance of the granular system as a function of the offset charge $q_i$,
wich appears in the partion function through the phase $\chi_i$. This dependence
is shown in Fig. \ref{gqi} for the same ${\tilde\alpha}$ value as in Fig. \ref{gt} at
a temperature $T = E_{\rm c}/10$. One finds a maximum of the conductance at
the degeneracy point ($q_i = 0.5$), similar to the well-known case of a single
junction in the presence of Coulomb blockade (solid line in Fig. \ref{gqi}). Again, we 
observe a decrease in the conductance for the granular system, as compared with
the single junction with the same effective parameters (${\tilde\alpha}, \epsilon$), 
in the absence of charge-offset ($q_i = 0$). However, close to the degeneracy point,
we find similar values for both systems, that in fact coincide for $q_i = 0.5$
within the error bars of our numerical results.

\begin{figure}
\includegraphics[width=18cm]{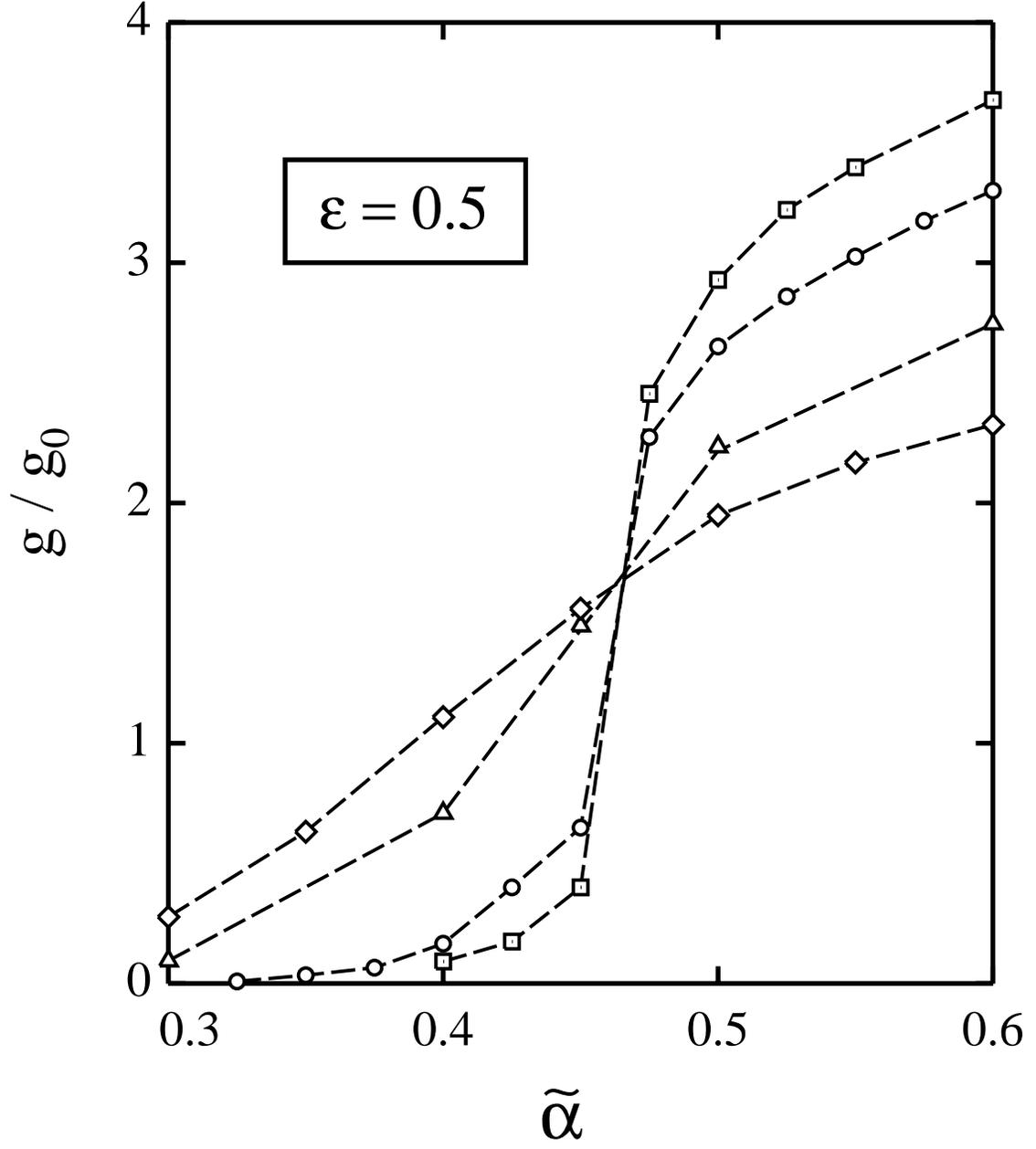}
\caption{\label{galpha}
 Normalized conductance as a function of ${\tilde\alpha}$ for a granular system 
with $\epsilon = 0.5$. Different symbols represent several temperatures:
$\beta \overline{E_{\rm c}} = 40$ (squares), 30 (circles), 20 (triangles), and 14 
(diamonds). } 
\end{figure}

The conductor-to-insulator phase transition at low temperatures in the parameter
space ($\epsilon, {\tilde\alpha}$) can be also analyzed fron MC simulations.
With this purpose, we have studied the dependence of the conductance on
${\tilde\alpha}$ and $T$ for several values of $\epsilon$. For a given
$\epsilon$ one expects a conducting regime for ${\tilde\alpha}$ larger than
some critical value ${\tilde\alpha}_c$ (see Fig. \ref{avse} above). In our numerical
analysis this means that we expect a crossover from a region in which 
$g$ increases (large ${\tilde\alpha}$) to another regime in which $g$ decreases 
(small ${\tilde\alpha}$) as $T$ is lowered. This is shown in Fig. \ref{galpha}, 
where we have displayed the dependence of the conductance $g$ on ${\tilde\alpha}$
for $\epsilon = 0.5$ at several temperatures. Each kind of symbols represents
MC results at a given temeprature, with $T$ increasing from top to bottom
in the right part of the Figure. From the crossing point of these curves we
can estimate a critical value ${\tilde\alpha}_c = 0.46 \pm 0.02$  for
$\epsilon = 0.5$.

\begin{figure}
\includegraphics[width=18cm]{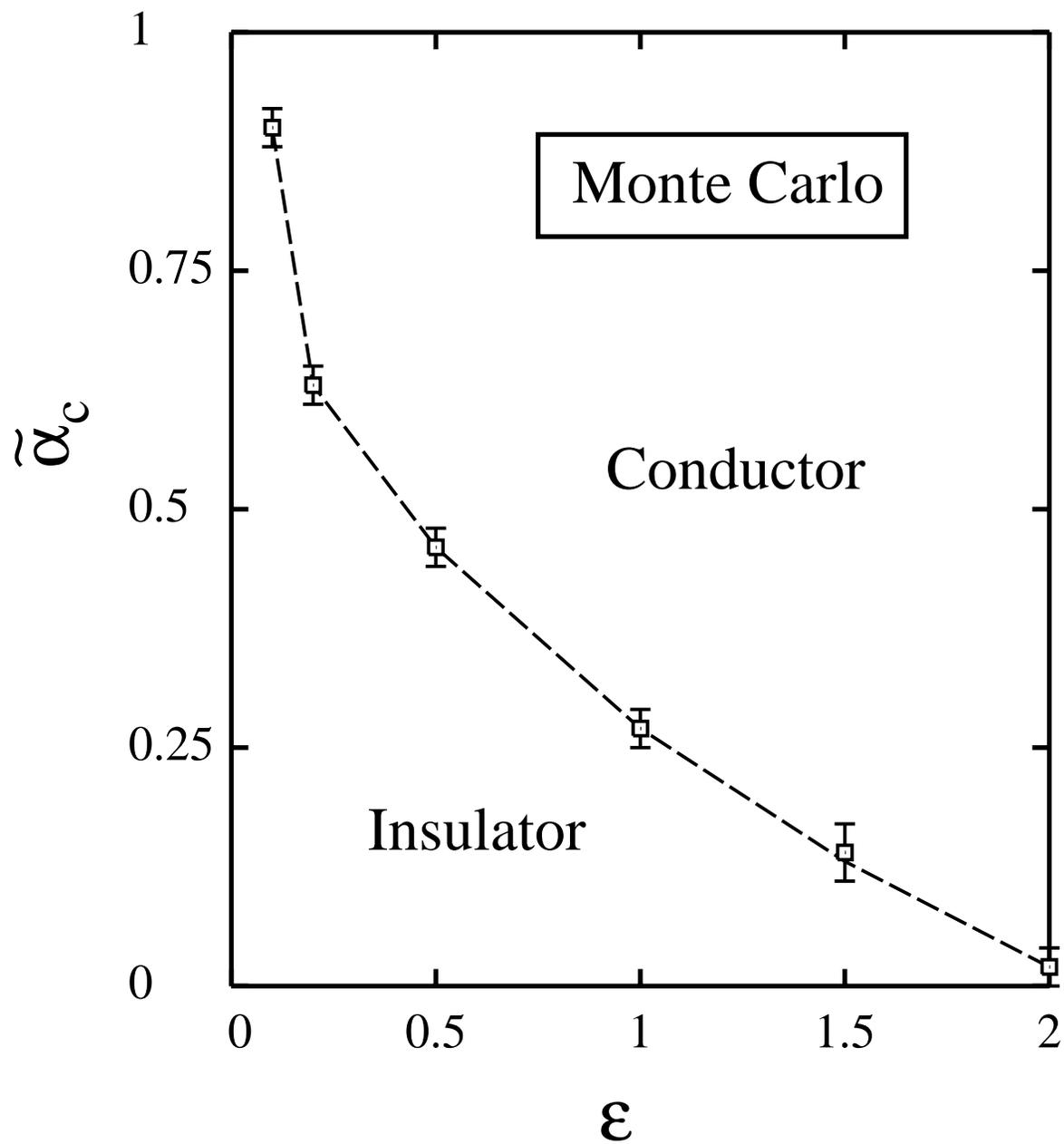}
\caption{\label{alphaeps}
 The critical value of the parameter ${\tilde\alpha}$ for the insulator-to-conductor
transition is plotted vs the parameter $\epsilon$.
Symbols were derived from Monte Carlo simulations, as indicated in the text. }
\end{figure}

In a similar way we have obtained ${\tilde\alpha}_c$ for other $\epsilon$
values, and the results are presented in Fig. \ref{alphaeps}. Symbols are data 
points derived from MC simulations, and the dashed line is a guide to the eye.

\section{Conclusions}
We have presented a method to analyze the conductance of metallic
granular systems in the range of parameters where Coulomb blockade
effects are important. The method is a natural extension of the techniques
widely used to study the competition between charging
effects and tunnel conductance in single junctions. 
The method allows us to include disorder, and becomes exact
when the average coordination number of a grain is large.
A related scheme, used to study ^^ ^^ grains " with a degenerate
level is presented in\cite{FG02}. This model is a natural
extension of the Hubbard model used to describe the transition between
a Mott insulator and a correlated metal, and does show this transition.
It would be interesting to compare the properties of this model
to the results presented here. A related problem is that of superconducting
grains coupled by the Josephson effect. This model, neglecting
the effect of normal quasiparticles, has been treated in\cite{AS02},
using the CPA method, which is the analog of the Dynamical
Mean Field Theory used here.

In the absence of non equilibrium effects, we find that the
low temperature behavior of the system is insulating. There
is a well defined Coulomb gap at low temperatures,  which,
when the intergrain conductances are large, can differ 
significantly from the intragrain charging energy. This
behavior is different from the low temperature conductance
of a single junction, where the gapless nature of the 
leads gives rise to a power law dependence of the conductance,
due to cotunneling processes. Our results are valid provided
that the temperature is much greater than the intradot level spacing.
Within this restriction, and in the absence of non equilibrium effects,
we find no evidence for a phase transition between an insulating
and a metallic phase\cite{ET02}. 

Non equilibrium effects can lead to a zero temperature phase transition,
in a similar way to the transition which occurs in a single junction.
It is interesting to note that this phase transition remains restricted
to zero temperature, even when an array of grains with large
coorination number is considered. This is so because the most favorable
case for a metallic regime is when the environment of a given dot
is itself metallic. This reduces the problem to that of a single
grain connected to metallic leads, which cannot have
a finite temperature phase transition, irrespective of the
coupling to the leads.

\section{Acknowledgements.}
We have benefited from useful conversations with S. Florens,
A. Goerges, G. Kotliar, and A. Zaikin.
We acknowledge financial support from CICyT (Spain)
through grant PB0875/96, CAM (Madrid) through grant 07N/0045/1998,
and the European Union through grant ERBFMRXCT960042.

~\newpage
 
\end{document}